# Terahertz channel performance under dynamic water surface reflections

Yapeng Ge, Jiacheng Liu, Jiayuan Cui, Mingxia Zhang, Wenbo Liu, Peian Li, Houjun Sun, Jianjun Ma

*Abstract*—As the terahertz (THz) band emerges as a pivotal technology for next-generation wireless communications, accurate channel modeling in dynamic environments becomes increasingly critical, particularly for scenarios involving reflective interactions with water surfaces. This article presents comprehensive experimental and theoretical investigations into THz channel (120-320 GHz) performance under dynamic water surface reflections. By developing and validating a modified dual-scale scattering model based on the improved integral equation model ($I^2EM$), this work systematically evaluates channel characteristics, such as signal power loss and bit error rate (BER), across various dynamic aquatic scenarios. Laboratory experiments and real-world natatorium measurements demonstrate the model's efficacy in capturing complex temporal and spatial scattering behaviors, offering vital insights and robust predictive capabilities essential for deploying practical THz communication systems in aquatic and sports environments.

## I. INTRODUCTION

As the demand for ultra-high-speed wireless communications intensifies, the terahertz (THz) band (0.1-10 THz) has garnered considerable attention as a foundational pillar of next-generation wireless communication systems due to its unprecedented spectral bandwidth and potential to support data rates in the terabit-per-second range [1, 2]. The unique properties of THz channels - narrow beamwidth, short wavelength, high directivity, and ultra-high frequency - make them ideal for integrated sensing and communication (ISAC) in dense, localized environments [3, 4]. In sports venues such as natatoriums, stadiums, and arenas, THz channels can support real-time data acquisition for applications like wearable-free athlete tracking, immersive broadcasting, and environment-aware beamforming [5, 6]. This positions THz-ISAC as a key enabler of next-generation smart sports infrastructure [7, 8]. THz channel also provides non-invasive and highly directional features [9], which are crucial in dense and human-occupied aquatic scenarios. Nevertheless, the deployment of THz systems in indoor settings is non-trivial due to its unique propagation characteristics - the short wavelengths and high frequencies exert channel performance degradation through both molecular absorption and surface scattering [10, 11]. THz channels are especially susceptible to environmental obstructions and multipath interference due to their limited diffraction and low penetration capabilities [12]. Consequently, accurate modeling of channel behavior, particularly with respect to interaction with material boundaries, is a prerequisite for robust system design.

To date, substantial research has been conducted on the reflection, scattering, and absorption characteristics of THz channels across common materials such as plaster walls, metals, and glass windows [13-15]. These studies have yielded a variety of surface scattering models that relate material microstructure and dielectric constant to incident angle, polarization, and frequency. Models such as the Kirchhoff Approximation, Integral Equation Model (IEM), and its advanced variants, as the Small Perturbation Method (SPM) and the Improved IEM ($I^2EM$), have been widely adopted to characterize surface-induced channel perturbations [16]. Extensions of these models have enabled scenario-specific THz channel characterization in indoor offices, vehicular communication networks, and smart buildings [17, 18]. However, these modeling efforts have predominantly focused on solid surfaces, whose geometry and material properties remain largely static during operation. In contrast, liquid surfaces, particularly water, introduce a fundamentally different scattering regime due to their spatiotemporal variability. When disturbed by external forces - such as swimmers, ventilation currents, or wave generators - water surfaces evolve into dynamic patterns comprising ripples and waveforms. These features cause time-varying perturbations in the local reflection geometry, altering both the amplitude and phase of the scattered wavefronts. In THz communication systems, which typically employ narrow-beam antennas, this translates into severe channel decorrelation and angular dispersion, even over short ranges [19].

Contemporary research on wireless communication over water has primarily focused on maritime or long-range

This work was supported in part by the National Natural Science Foundation of China under Grant.
Corresponding author: *Jianjun Ma  (Beijing Institute of Technology).*

riverine scenarios at sub-6 GHz and millimeter-wave (mm-wave) bands. These include power delay profile analyses [20], diversity gain studies [21], evaporation duct modeling [22], and statistical fading model [23, 24]. Specific attention has also been paid to over-sea [25], inland river [26], and urban water-land propagation scenarios [27]. Most of these models emphasize deterministic ray-tracing or empirical curve fitting at lower frequencies (typically < 30 GHz), and are validated under quasi-static or long-range conditions. The limited attention to the THz regime stems from both instrumentation constraints and the lack of suitable models for capturing small-scale surface roughness superimposed on large-scale waves. A recent research demonstrated the use of modified parabolic equation models to predict THz channel performance over evaporation ducts [28]. Similarly, Liao *et al.* [27] modeled mm-wave line-of-sight (LOS) and non-line-of-sight (NLOS) water reflection in pond-like urban setups at 28 GHz. However, as summarized in Table I, present approaches either neglect the high spatial-frequency components of water ripples or are restricted to fixed-path geometries with low angular resolution, which can lead to insfficient channel predictions and system performance evaluations.

In this work, we propose a modified dual-scale scattering model based on the improved integral equation model ($I^2EM$), tailored for characterizing THz channel interactions with dynamic water surfaces. Our approach jointly captures large-scale sinusoidal waveforms (macroscopic geometry) and small-scale roughness (stochastic perturbations), providing an analytically tractable and physically accurate model for bistatic THz channel scattering from disturbed liquid interfaces. This is of particular significance in natatoriums, where THz channels often reflect off water surfaces under grazing angles and fluctuating curvature induced by swimmer activity. To validate the model, we conducted a two-stage experimental campaign. The first involved laboratory-scale, repeatable wave generation under controlled frequency and angular conditions. The second was a realistic natatorium deployment, where near-grazing reflections from swimmer-induced waveforms were measured under LOS and NLOS configurations. Moreover, bit error rate (BER) simulations under various modulation schemes were conducted.

The following of this article is organized as: Section II presents the laboratory measurement setup and results, including the experimental configuration for simulating dynamic water surfaces, the dual-scale scattering model based on the $I^2EM$, and the analysis of power loss characteristics across different frequencies, angles, and wave conditions. It also includes a detailed power profile performance evaluation via cumulative distribution function (CDF) analysis. Section III extends the research to real-world aquatic environments through a swimming pool measurement campaign, covering NLOS experimental configurations, power loss behaviors under natural wave states, statistical CDF analyses, and BER simulations. Section IV summarizes the research findings and emphasizes the significance of the study.

## II. LABORATORY MEASUREMENT AND RESULTS

### A. Experimental setup

In order to investigate the channel scattering behavior induced by dynamic water surfaces, an indoor laboratory measurement system was designed to emulate repeatable, controlled channel conditions and to characterize frequency- and angle- dependent reflection losses. The experimental configuration is schematically illustrated in Fig. 1. A rectangular water tank measuring 120 cm in length and 35 cm in width was employed as the physical medium for wave generation. A mechanically actuated oscillating paddle, matching the full width of the tank, was mounted at one end and configured to pivot periodically around its bottom edge. This mechanism generated one-dimensional sinusoidal surface waves along the longitudinal axis of the tank, thereby enabling precise control over wave periodicity and amplitude. By adjusting the swing frequency and displacement angle of the paddle, surface profiles with variable steepness and spatial wavelength could be synthesized, closely mimicking real-world aquatic perturbations. Three discrete wave conditions were synthesized in the tank to represent increasingly complex surface states - *calm surface* (WI1) without wave excitation; *low-amplitude periodic waves* (WI2), with a crest-to-trough height of approximately 2 cm; *high-amplitude periodic waves* (WI3), reaching 4 cm in height. These conditions were selected to emulate swimmer-induced or mechanically disturbed surface dynamics in aquatic facilities. The wave height was measured using a piezoelectric wave sensor (Yunjing Tianhe TH-LGZ1) placed near the reflection point, operating at a sampling rate of 1 kHz to capture real-time surface dynamics without obstructing the THz beam path. Throughout the experimental conditions, the generation of surface waves remained within the linear regime, with minimal occurrences of wave breaking; thus, the resultant surface roughness was chosen to be ~ 100 nm in root-mean-square (RMS) height [31] for reflection modeling purposes throughout this study.

TABLE I
SUMMARY OF THE OVER-WATER CHANNEL PERFORMANCE RESEARCH WORKS

| Frequency | NLOS/LOS | Model Method | Research Scenario | Ref. |
|---|---|---|---|---|
| 1.9 GHz | Both | Rician & Rayleigh PDP models | Aegean Sea marine | [20] |
| 3.15 GHz - 12.55 GHz | LOS | Naval Postgraduate School (NPS) and Advanced Propagation (APM) duct models | South China Sea evaporation duct | [22] |
| 1.39 GHz, 4.5 GHz | Both | Improved two-ray model with Gaussian randoms | Ship-to-shore | [24] |
| 1.5 GHz, 3 GHz | Both | Two-way parabolic equation (2WPE) with Pade(2,2) approximation | Rough sea (simulated) | [29] |
| 1 GHz | LOS | Dual-path model and modified Longley-Rice model | Buoy-to-tower | [30] |
| 28 GHz | Both | Path loss model with water surface reflection and building diffraction | Urban water–land (campus pond) | [27] |
| 0.5 GHz - 5 GHz | NLOS | Proposed path loss model based on ABG model | Complex water surface (obstacles) | [23] |
| 5.9 GHz | LOS | Ray tracing simulation with electromagnetic parameter calibration | Inland river / ship-to-shore | [26] |
| 0.14 THz | LOS | Modified PE model with evaporation duct and atmospheric absorption | Over-sea THz link | [28] |

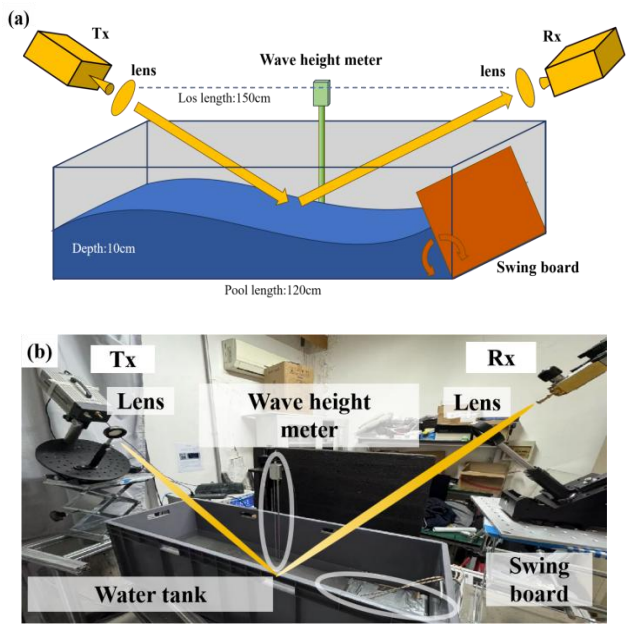

**Figure 1** Experimental setup configuration in laboratory. (a) Schematic diagram and (b) actual experimental setup.

To evaluate the spectral sensitivity of the water surface reflection, the system operated across six frequencies: 120 GHz, 140 GHz, 160 GHz, 220 GHz, 270 GHz, and 320 GHz. These frequencies span both lower and upper portions of the D-band and G-band, capturing key transition regions in frequency-dependent absorption and scattering behavior. For the lower band (120-160 GHz), a Ceyear 1465D vector signal generator produced baseband signals up to 20 GHz, which were up-converted using a 12× frequency multiplier (Ceyear 82406B). Signal radiation and reception employed HD-1400SGAH25 horn antennas with dielectric lenses. The received signal was measured using a Ceyear 71718 power sensor with a dynamic range suitable for capturing subtle variations in reflected power. For the higher band (220-320 GHz), the same baseband source was used in conjunction with an 18× multiplier (Ceyear 82406D). Radiation and reception were carried out using Ceyear 89901S horn antennas, chosen for their optimized gain and reduced side-lobe levels in the higher frequency range. All received signals were routed to a Ceyear 2438PA power meter, which interfaces with a dedicated computer system for real-time data acquisition. The power meter's 50 ms response time ensures accurate integration over short-term fluctuations in wave reflection. A sampling rate of 7 Hz was selected to synchronize with the ~1 s wave cycle period, ensuring adequate temporal resolution to resolve intra-cycle variations while maintaining manageable data volume. Due to the physical constraints of the Tx and Rx mounting structure, only vertically polarized signals could be transmitted and received. Collectively, this measurement platform enables rigorous characterization of frequency-selective, angle-dependent, and wave-state-sensitive THz channel reflection phenomena over dynamic liquid surfaces. The setup balances experimental repeatability with sufficient fidelity to support validation of theoretical scattering models in time-varying aquatic environments.

The measurement system was configured in a bistatic reflection geometry, as in Fig. 1(a). The transmitter (Tx) and receiver (Rx) units were symmetrically placed on opposite sides of the tank, each 75 cm from the tank's midpoint. Both units were equipped with 10 cm focal length dielectric lenses mounted on horn antennas to ensure narrow-beam, spatially coherent illumination of the water surface. The antennas were mechanically aligned such that the incident and reflected beams converged at a specular reflection point located at the geometric center of the tank, enabling controlled variation of incidence angles without altering lateral separation between Tx and Rx.

To achieve different incidence configurations (notably 45° and 60°, due to the limitations of the experimental setup), the vertical height and inclination angle of the Tx and Rx were adjusted synchronously, ensuring that the reflected path remained aligned with the expected specular geometry. Importantly, the horizontal baseline between the antennas was maintained constant to isolate angular effects. A wave height sensor was positioned adjacent to the central reflection point to monitor wave amplitude in real time, thereby facilitating correlation between surface conditions and observed signal fluctuations.

### B. Theoretical Model

For calm surface (WI1) - where the water surface can be approximated as planar - the reflection behavior of THz channel is governed by classical Fresnel theory. This provides a baseline for specular reflectivity, decomposed into vertical and horizontal polarizations. The Fresnel coefficients are computed as functions of the complex permittivity of water and the angle of incidence [32], and are expressed as

$$R_v = \frac{\varepsilon_r \cos\theta - \sqrt{\mu_r \varepsilon_r - \sin^2\theta}}{\varepsilon_r \cos\theta + \sqrt{\mu_r \varepsilon_r - \sin^2\theta}}$$
$$R_h = \frac{\mu_r \cos\theta - \sqrt{\mu_r \varepsilon_r - \sin^2\theta}}{\mu_r \cos\theta + \sqrt{\mu_r \varepsilon_r - \sin^2\theta}} \quad (1)$$

where $\mu_r$ and $\varepsilon_r$ are the relative permeability and permittivity of water, respectively, and $\theta$ is the incidence angle. These coefficients serve as a critical reference for benchmarking measurements under static surface conditions. The parameter, $\varepsilon_r$ was calculated using the double-Debye dielectric model [4] for pure water at the corresponding ambient temperature (20°C for laboratory tests and 25°C for natatorium measurements), ensuring frequency-specific accuracy in both the real and imaginary parts of permittivity. However, in the presence of water surface waves (WI2, WI3), the water interface exhibits a non-planar profile that must be modeled beyond Fresnel optics. To address the scattering effects of dynamic water surfaces with sinusoidal waves and superimposed small-scale roughness, we choose the I²EM model, due to its superior capability to handle surface scattering from moderately rough and complex geometries [13, 33], and extend it. This new model captures the interplay between large-scale wave geometry and small-scale roughness, which significantly influences THz channel reflection. This extension accounts for both deterministic waveforms on a large scale and statistically characterized roughness on a small scale [34]. The surface geometry is modeled as a periodic sinusoid defined by the height function, as

$$z_{\text{large}}(x,y) = A \cos\left(\frac{2\pi}{G_{mm}} y\right) \quad (2)$$

where $A$ is the wave amplitude and $G_{mm}$ is the spatial period. Superimposed on this waveform is a fine-scale roughness characterized by its root-mean-square (RMS) height $\sigma$ and correlation length $l$, assumed to follow an isotropic Gaussian autocorrelation function. Within this framework, the reflected field is decomposed into a coherent specular component and an incoherent diffuse scattering term [13]. The coherent reflectivity, for either polarization $R_{v/h}$, is modulated by an exponential factor representing phase decoherence due to roughness, as

$$R_v = \text{ref}_v^{\text{diff}} + e^{-k^2 \sigma^2 \cos^2\theta} \cdot |r_v^{\text{spec}}|^2$$
$$R_h = \text{ref}_h^{\text{diff}} + e^{-k^2 \sigma^2 \cos^2\theta} \cdot |r_h^{\text{spec}}|^2 \quad (3)$$

Here, $R_v$ and $R_h$ represent the vertically and horizontally polarized reflectivity, respectively. The terms $\text{ref}_v^{\text{diff}}$ and $\text{ref}_h^{\text{diff}}$ denote the diffuse reflectivity components, while $r_v^{\text{spec}}$ and $r_h^{\text{spec}}$ correspond to the Fresnel-based specular reflection coefficients. The parameter $k$ is the wavenumber ($k = 2\pi/\lambda$, with $\lambda$ as the wavelength). The exponential term accounts for roughness-induced coherence loss in the specular component, which reduces for larger incidence angles [19]. This formulation ensures that increasing surface roughness induces a reduction in coherent reflectivity, consistent with physical intuition and experimental observations.

The calculation of the diffuse reflection term depends on the scattering coefficients. These coefficients form the basis of the single-scatter surface scattering model. For co-polarized (pp) cases in bistatic setups with simplified Green's function [35], the equation is written as

$$\sigma_{qp}^s = S(\theta, \theta_s) \frac{k^2}{2} \exp\left[-\sigma^2 \left(k_z^2 + k_{z_s}^2\right)\right]$$
$$\cdot \sum_{n=1}^{\infty} \sigma^{2n} |I_{qp}^n|^2 \frac{W^{(n)}(k_{x_s} - k_x, k_{y_s} - k_y)}{n!} \quad (4)$$

The components $k_x$, $k_y$, and $k_z$ are derived from the incident direction angles $\theta$ (elevation) and $\Phi$ (azimuth), expressed as: $k_x = k\sin\theta\cos\phi$, $k_y = k\sin\theta\sin\phi$, $k_z = k\cos\theta$. Similarly, the scattered direction angles $\theta_s$ and $\Phi_s$ define the components $k_{xs}$, $k_{ys}$, and $k_{zs}$. The $W^{(n)}$ corresponds to the Fourier transform of the n-th power of the surface correlation function. Additionally, $S(\theta, \theta_s)$ refers to the bistatic shadowing function. The $I_{qp}^n$ under the complete form of the Green's function is given as

$$I_{qp}^n = (k_z + k_{z_s})^n f_{qp} \exp(-\sigma^2 k_z k_{z_s}) +$$
$$\frac{1}{4}[(k_{z_s} - q)^n F_{ppup} \exp\{-\sigma^2 [q^2 - q(k_{z_s} - k_z)]\}]|_{u,v=-k_x,y}$$
$$+(k_{z_s} + q)^n F_{ppdn} \exp\{-\sigma^2 [q^2 + q(k_{z_s} - k_z)]\}]|_{u,v=-k_x,y} \quad (5)$$
$$+(k_z + q)^n F_{ppups} \exp\{-\sigma^2 [q^2 - q(k_{z_s} - k_z)]\}]|_{u,v=-k_{xs},y_s}$$
$$+(k_z - q)^n F_{ppdns} \exp\{-\sigma^2 [q^2 + q(k_{z_s} - k_z)]\}]|_{u,v=-k_{xs},y_s}]$$

$q = \sqrt{k^2 - u^2 - v^2}$ and the field coefficients ($f_{qp}$, $F_{qp}$) are given in literature [16].

To reconcile the micro-scale scattering with macro-scale wave curvature, a coordinate system transformation is required. Each surface element along the sinusoidal wave profile is treated as a locally planar patch with an associated local incidence angle and polarization basis. The local coordinate system is defined by the local polarization vectors derived from the local incidence angle, whereas the global coordinate system is determined by the global polarization vectors. The relationship between the global reflectivity ($R_{vss}$, $R_{hss}$) and the local reflectivity ($R_{vs}$, $R_{hs}$) is governed by the equations

$$R_{vss} = (v \cdot h_{pr})^2 \cdot R_{hs} + (v \cdot v_{pr})^2 \cdot R_{vs}$$
$$R_{hss} = (h \cdot h_{pr})^2 \cdot R_{hs} + (h \cdot v_{pr})^2 \cdot R_{vs} \quad (6)$$

where $v$ and $h$ denote the vertical and horizontal polarization vectors in the global coordinate system, respectively, while $v_{pr}$

and $h_{pr}$ represent their counterparts in the local coordinate system. The dot products quantify the projection alignment between polarization vectors across scales, ensuring physically consistent reflectivity transitions. The final step involves spatial integration of both coherent and diffuse reflectivity over a single spatial period of the sinusoidal wave, as

$$R_v = \frac{\int R_{v_{ss}}(y)dy}{G_{mm}} + R_{v,multi} \cdot \rho_s$$
$$R_h = \frac{\int R_{h_{ss}}(y)dy}{G_{mm}} + R_{h,multi} \cdot \rho_s \quad (7)$$

where the first term integrates the global reflectivity and then normalizes it, and the second term takes into account the secondary scattering. $R_{v/h,multi}$ is the secondary reflectivity, and $\rho_s$ is the ratio of unobstructed secondary scattering points. We found that the contribution of secondary scattering is almost negligible, as there is no geometric shadowing by the waves with strength in this work, which will be demonstrated later. This dual-scale formulation yields frequency- and angle-resolved reflectivity patterns that account for the cumulative impact of macro-waveform geometry and micro-scale perturbations.

### C. Measurement and theoretical results

The experimental data and theoretical predictions are presented in Fig. 2, which collectively offer a validation of the dual-scale scattering model developed for characterizing THz channel interactions with dynamic water surfaces. The measurements span a range of frequencies (120-320 GHz), water wave amplitudes (0-4 cm), and incidence angles (45° and 60°), enabling detailed insight into frequency-dependent attenuation mechanisms and angular scattering dynamics under varying hydrodynamic regimes.

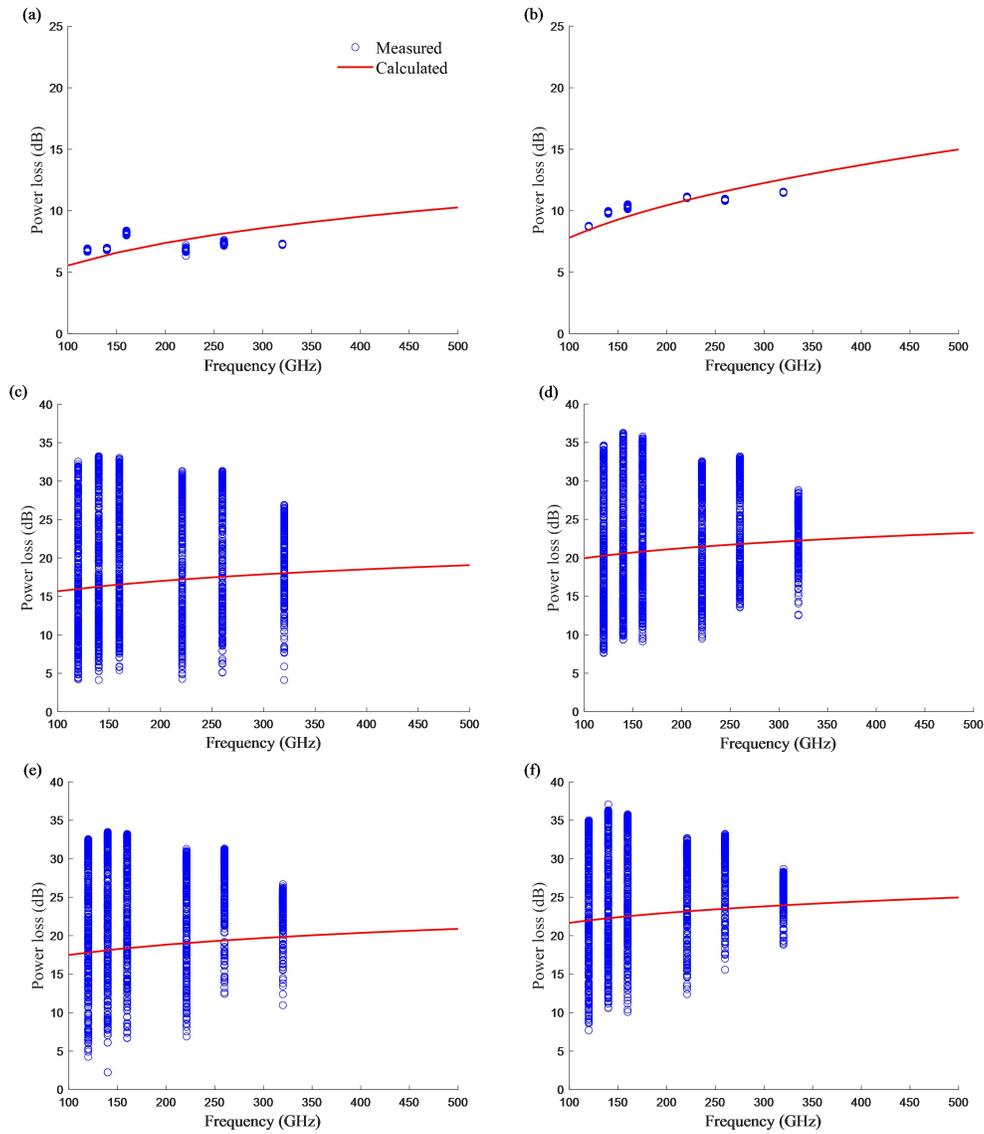

**Figure 2** Laboratory experimental data and theoretical predictions. (a), (c) and (e) correspond to *calm surface* (WI1) without wave excitation, *low-amplitude periodic waves* (WI2) with a crest-to-trough height of approximately 2 cm; *high-amplitude periodic waves* (WI3) reaching 4 cm in height, at an incidence angle of 45°, and (b), (d) and (f) correspond to wave WI1, WI2 and WI3 at an incidence angle of 60°, respectively.

Under calm surface conditions - corresponding to the baseline cases WI1 at both 45° and 60° incidence angles (Fig. 2(a) and 2(b)) - the measured power loss exhibits remarkable consistency and low dispersion across all frequencies. This stability is attributed to the absence of water wave-induced perturbations, allowing specular reflection to dominate the received signal. In this regime, the theoretical model, grounded in Fresnel reflectivity, demonstrates good agreement with measured data. A slight increase in reflection loss with frequency is observed, which arises from the enhanced absorption by the water surface, even though both the real and imaginary parts of water's complex permittivity decrease at higher THz frequencies [36, 37]. Furthermore, consistent with the electromagnetic boundary conditions at the air-water interface, power loss is marginally higher at 60° incidence than at 45° (around 2.2-4.7 dB higher), due to greater susceptibility to angular deviation and increased interaction with the lossy dielectric medium. This phenomenon occurs only when the incidence angle is smaller than the Brewster angle [38], which is 74° at 120 GHz, 73° at 140 GHz and 160 GHz, 71° at 220 GHz and 230 GHz. The influence of incidence angle on reflection loss is evident across all surface conditions, after a broad comparison of the 6 plots in Fig. 2. Notably, the difference in power loss between these two angles remains approximately 3-4 dB for all three surface states (WI1, WI2, WI3). This consistent gap suggests that the geometric shadowing from surface slopes within this angular range is negligible. The surface features remain within the regime where all local facets are still visible to the incident wavefront.

As the water surface transitions into dynamic states with increasing wave amplitude (WI2 and WI3), shown in Fig. 2(c)-(f), the measured power loss distributions become significantly more dispersed. This broadening of data points reflects the stochastic nature of surface-induced perturbations, which modulate both the local angle of incidence and the surface orientation across the illuminated footprint. Importantly, although the variance of individual measurements increases under dynamic conditions, the central trend of the measured data aligns well with the theoretical predictions, represented by the smooth red curves. This agreement indicates that the model successfully captures the average scattering behavior by integrating over the sinusoidal macrostructure and incorporating roughness-induced phase decoherence at smaller scales. It should be noted that the validity of the dual-scale I$^2$EM model relies on foundational assumptions, namely that the surface wave behavior remains within the linear regime and that secondary scattering effects are negligible. These conditions are well-satisfied under the controlled laboratory settings of our study, where wave amplitudes are modest and no wave breaking occurs.

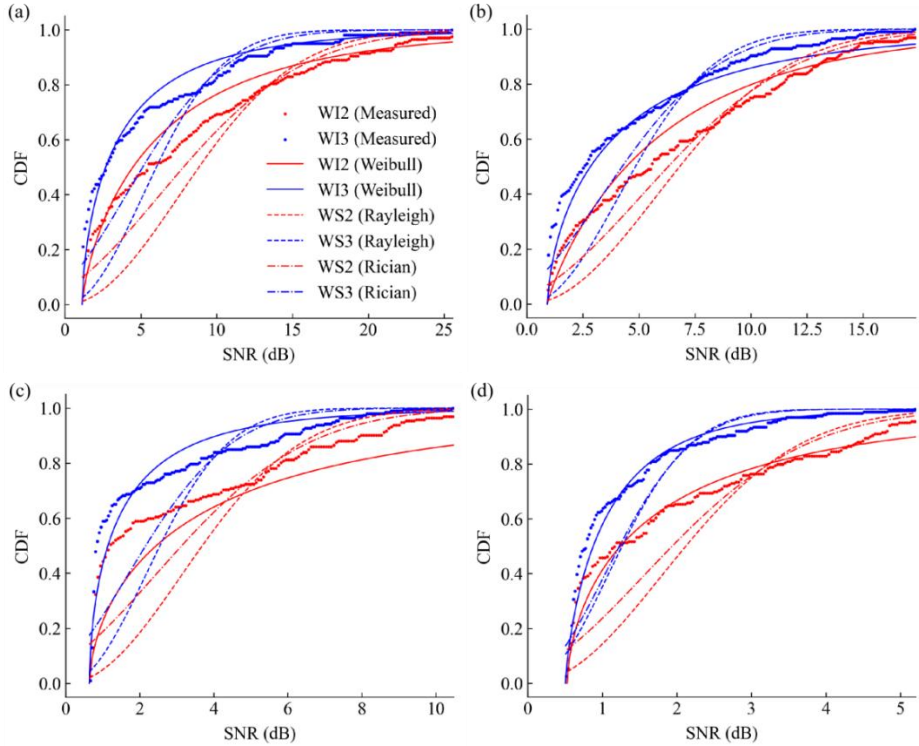

**Figure 3** CDF analysis of experimental results. The experimental conditions are as (a) 45° at 140 GHz (b) 60° at 140 GHz (c) 45° at 220 GHz (d) 60° at 220 GHz. (b), (c) and (d) keeps the same legend with (a).

TABLE II
COMPARISON OF FITTING GAP AND FLUCTUATION ERROR (FE) FOR DIFFERENT FADING DISTRIBUTIONS

| Incident angle | Wave Type | Fitting Gap | | | Fluctuation Error | | |
|---|---|---|---|---|---|---|---|
| | | Rayleigh | Rician | Weibull | Rayleigh | Rician | Weibull |
| 45° | WI1 | 1.0819 | 0.0857 | 0.0058 | 0.1708 | 0.0410 | 0.0196 |
| | WI2 | 0.5489 | 0.5489 | 0.1009 | 0.1011 | 0.1011 | 0.0303 |
| | WI3 | 0.8308 | 0.8191 | 0.199 | 0.1155 | 0.1125 | 0.0410 |
| 60° | WI1 | 1.0912 | 0.4203 | 0.0024 | 0.1774 | 0.1077 | 0.0113 |
| | WI2 | 0.2285 | 0.5750 | 0.0797 | 0.0716 | 0.0715 | 0.0349 |
| | WI3 | 0.4499 | 0.2285 | 0.0653 | 0.0973 | 0.0897 | 0.0365 |

Notably, increasing the water wave amplitude from 2 cm (WI2) to 4 cm (WI3) results in a systematic elevation of mean power loss across all frequencies and both angles. Physically, higher water wave heights introduce steeper local slopes and deeper troughs, which exacerbate bistatic scattering and diminish coherent specular reflection. As such, in Fig. 2(e) and (f) (WI3), the observed power loss is approximately 10-15 dB higher than in the calm cases (WI1), emphasizing the critical role of water wave-induced scattering in degrading channel behavior.

Furthermore, the trend of increasing power loss with frequency is consistently observed across all water states and angles. This monotonic increase is due to both the intrinsic absorption loss of water and the enhanced scattering efficiency of shorter wavelengths, which are more sensitive to sub-wavelength roughness. Consequently, the higher-frequency bands (270-320 GHz) exhibit greater channel degradation, particularly under wave conditions.

*D. Power profile performance*

The statistical characterization of received THz channel power under varying water surface conditions was analyzed through cumulative distribution function (CDF) evaluations, as depicted in Fig. 3. These results, obtained across two representative frequencies (140 GHz and 220 GHz) and two incidence angles (45° and 60°), enable a quantitative understanding of the stochastic fading phenomena induced by surface dynamics. Under all configurations, a clear trend emerges: signal-to-noise ratio (SNR) degrades as water wave amplitude increases. For both frequencies and angles, *calm water surfaces* (WI1) produce the highest SNR values, followed by *low-amplitude periodic waves* (WI2), with *large-amplitude periodic waves* (WI3) yielding the lowest SNR. This degradation is driven primarily by the progressive increase in wave amplitude, which amplifies the randomness of local surface orientation and the severity of multipath interference, thereby undermining the coherence of specular reflections.

Another notable observation is that SNR distributions are broader and statistically more dispersed at lower frequencies (140 GHz) compared to 220 GHz, particularly under dynamic water surface conditions. While we have observed higher frequencies to experience greater loss due to their increased sensitivity to surface perturbations, the narrower channel beams at higher frequencies lead to more deterministic reflections when local slope alignment is maintained, hence producing slightly less spread in the SNR's CDF. This observation underscores the complexity of THz channel propagation over time-varying medium and suggests that frequency selection must balance between angular sensitivity and multipath resilience.

At a given frequency, the incidence angle also plays a significant role in shaping the SNR distribution. Fig. 3(a)-(d) consistently show that the 60° incidence angle results in marginally lower SNR compared to 45°, especially under WI1 and WI2 scenarios. However, the performance gap narrows under WI3 conditions due to the increasing stochasticity of surface reflection geometry, which diminishes the relative advantage of smaller incidence angles. This highlights the interplay between geometric optics and statistical scattering in determining optimal beam alignment strategies for dynamic aquatic environments.

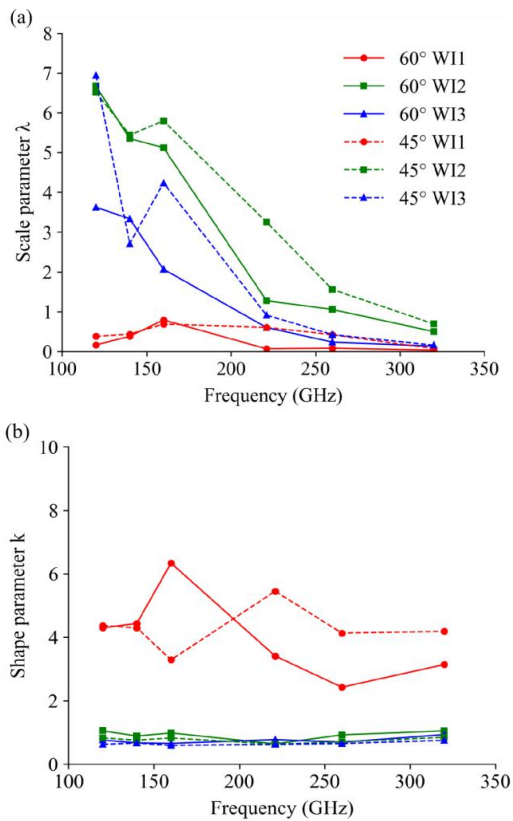

**Figure 4** Variation of Weibull parameters with frequency under different incident angles and wave heights. (a) Scale parameter $\lambda$; (b) Shape parameter k. (b) keeps the same legend with (a).

To provide deeper insight into the statistical nature of signal fading, three commonly adopted probabilistic models - Rayleigh, Rician, and Weibull - were fitted to the empirical CDFs as in Fig. 3. The comparative results, summarized in Table II, clearly demonstrate the superior performance of the Weibull distribution across all wave heights and angles. The

Weibull model consistently yields the lowest fitting gap and fluctuation error, indicating its exceptional capability in capturing both central tendency and tail behavior of the observed fading statistics [39]. This suggests that the THz channel over dynamic water surfaces exhibits compound fading characteristics - neither purely random (as in Rayleigh) nor solely dominant-path (as in Rician) - but rather a flexible distribution with heavy-tailed, amplitude-dependent characteristics. The effectiveness of the Weibull model is primarily due to its shape and scale parameters, which allow it to accommodate a broad range of fading severities and temporal variability, making it especially suitable for environments with hybrid deterministic-stochastic scattering properties such as indoor pools.

Figure 4 further supports this statistical assessment by illustrating how the Weibull parameters vary across frequency, angle, and wave height. The scale parameter ($\lambda$), which reflects the spread or average power level of the distribution, exhibits a distinct downward trend with increasing frequency. This is consistent with the frequency-dependent increase in path loss observed earlier and confirms that higher-frequency THz channels are more susceptible to power loss due to their stronger interaction with micro-scale surface structures. Moreover, $\lambda$ increases with wave amplitude, demonstrating that larger waves induce greater signal variability due to intensified surface perturbation. Interestingly, $\lambda$ values are generally higher at 45° incidence than at 60°, implying a larger dynamic range of received power when incident waves interact with less oblique angles, likely due to broader angular acceptance of reflected power under mild slopes. The shape parameter ($k$), on the other hand, governs the tail behavior of the distribution and provides a measure of the randomness and severity of the fading. As shown in Fig. 4(b), the $k$ values are highest under WI1 conditions and decline as wave amplitude increases. This decrease indicates a transition from concentrated signal power distributions toward more dispersed and heavy-tailed fading behavior, corroborating the observed degradation in link stability under dynamic surface conditions in Fig. 2. Unlike $\lambda$, the shape parameter does not exhibit a strong dependence on frequency, suggesting that wave-induced variability is the dominant factor shaping the statistical structure of the fading envelope rather than frequency-dependent propagation mechanisms.

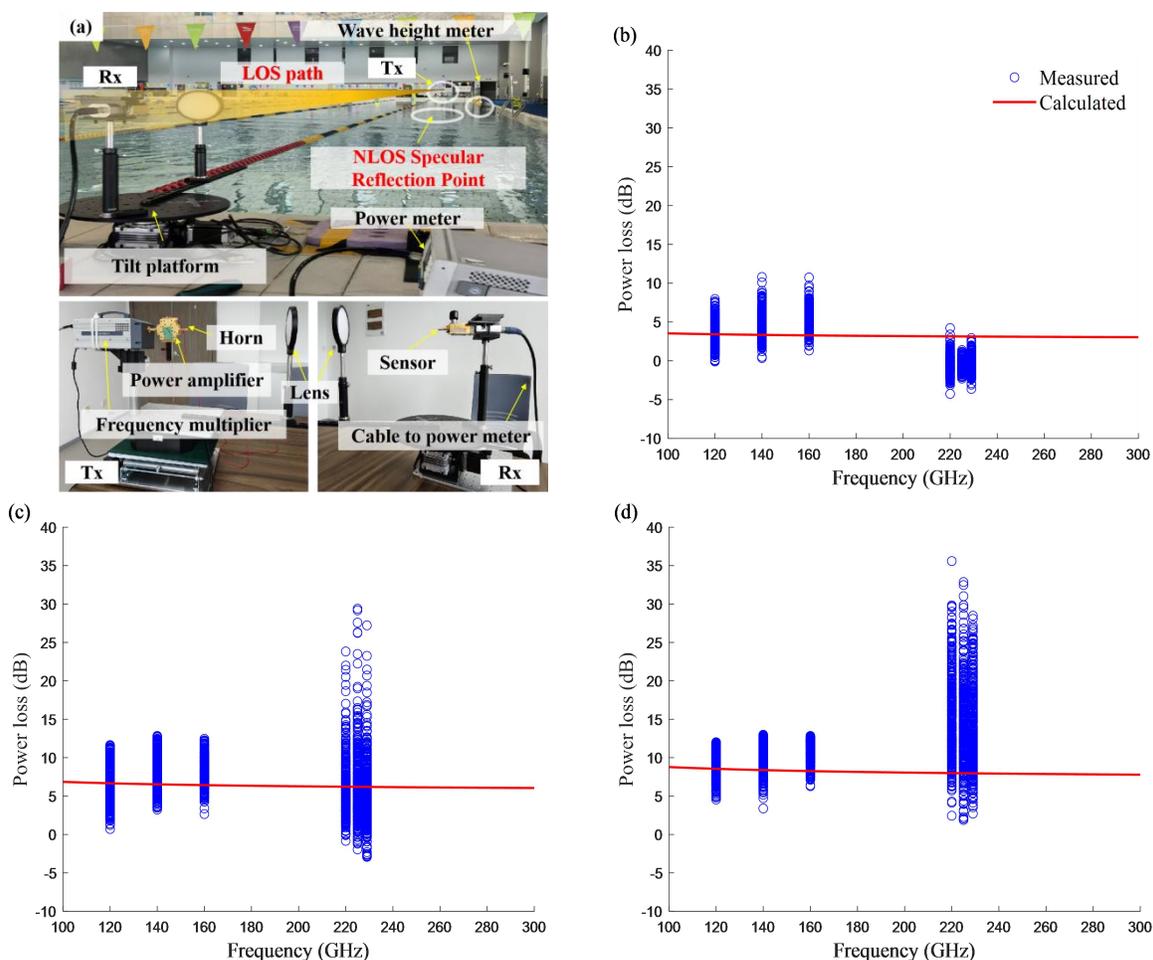

**Figure 5** Experimental setup, measured data and theoretical predictions. (a) Experimental setup configuration in swimming pool. Swimming pool experimental data and theoretical predictions with (b), (c) and (d) correspond to wave types of wave WS1, WS2 and WS3, respectively. (c) and (d) keeps the same legend with (b).

## III. Swimming pool measurement and results

To extend the applicability of the theoretical and laboratory-validated scattering model to real-world environments, a field measurement campaign was conducted in a standard 50-meter indoor swimming pool, in Liangxiang Campus at Beijing Institute of Technology. The purpose of this campaign was to evaluate the performance of THz channel under more realistic, large-scale aquatic conditions that feature natural wave propagation, ambient turbulence, and potential human-induced surface perturbations.

### A. Experimental setup configuration

The experiment was structured around two principal transmission scenarios: LOS communication, where the transmitted beam follows a direct channel path through air, and NLOS communication, where the channel is intentionally reflected off the water surface. The focus was placed on the NLOS configuration, which is of primary interest for real-world deployments in aquatic environments, such as swimmer monitoring or underwater relay systems, where direct paths may be blocked or geometrically constrained. As shown schematically in Fig. 5(a), the transmitter (Tx) and receiver (Rx) were mounted at opposite ends of the swimming pool, both positioned at a vertical height of 0.73 meters above the water surface. Each was equipped with a high-gain horn antenna and a dielectric lens of 15 cm focal length to ensure beam collimation and tight angular confinement. For the NLOS configuration, the antennas were precisely tilted downward such that their beams converged at the central specular reflection point on the water surface. This reflection geometry, established through geometric optics principles, resulted in a depression angle of 1.64° for both the Tx and Rx, ensuring symmetry and minimizing angular deviation errors. This low-angle configuration closely approximates grazing incidence, a regime in which THz beams exhibit high directionality but are highly sensitive to surface roughness and wave geometry.

To monitor the real-time water surface profile without interfering with the beam path, a non-intrusive wave height meter was placed adjacent to the specular reflection point. This instrumentation enabled correlation between instantaneous wave elevation and received signal fluctuations, allowing precise identification of dynamic reflection-induced channel degradation. The THz channel encountered during measurement consisted of three distinct conditions: a *calm surface* with slight ripple ~ 1 cm (WS1), *low-amplitude periodic waves* with ~4 cm crest-to-trough height (WS2), and *high-amplitude waves* reaching ~7 cm (WS3). These conditions captured typical operational environments in competitive or recreational pools, where swimmers, ventilation, or water displacement may induce spatiotemporal disturbances. The wave height was still recorded using a piezoelectric sensor (Yunjing Tianhe TH-LGZ1) operating at a 1 kHz sampling rate and the RMS roughness of approximately 100 nm was also adopted based on literature-reported values [31].

Measurements were performed across six frequencies: 120 GHz, 140 GHz, 160 GHz, 220 GHz, 225 GHz, and 230 GHz. For the lower three frequencies, the same equipment chain used in laboratory measurements was applied directly, ensuring consistency in hardware response and calibration. For the higher frequencies, a dedicated power amplifier

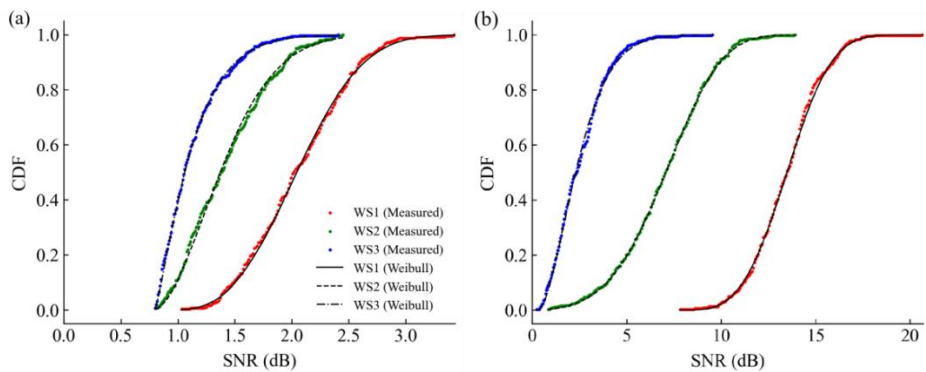

**Figure 6** CDF analysis for swimming pool experimental result. The experimental conditions are as follows: (a) NLOS 140 GHz (b) NLOS 220 GHz. (b) keeps the same legend with (a).

TABLE III

Comparison of Fitting Gap and Fluctuation Error (FE) for different distributions in swimming pool

| Wave Type | Fitting Gap | | | Fluctuation Error | | |
|---|---|---|---|---|---|---|
| | Rayleigh | Rician | Weibull | Rayleigh | Rician | Weibull |
| WS1 | 0.3930 | 0.0012 | 0.0011 | 0.1151 | 0.0095 | 0.0087 |
| WS2 | 0.0989 | 0.0045 | 0.0016 | 0.0583 | 0.0128 | 0.0071 |
| WS3 | 0.1093 | 0.0167 | 0.0021 | 0.0455 | 0.0217 | 0.0080 |

(TCNPA-220, operating at 220 - 230 GHz) was introduced at the transmitter to overcome increased free-space and reflection path loss (more detailed in reference [40]), especially in the presence of surface scattering and angular deflection. The use of high-frequency amplification was crucial to ensure sufficient dynamic range in the power measurements, particularly under WS2 and WS3 conditions where diffuse scattering dominates.

*B. Experimental and theoretical results*

The measurement results obtained from the 50-meter indoor swimming pool, shown in Fig. 5(b)-(d), provide a realistic validation of the proposed dual-scale scattering model under field conditions. The power loss is plotted against frequency for three distinct water surface scenarios: WS1, WS2, and WS3, representing increasing degrees of hydrodynamic complexity. The dominant wave periods for these scenarios are 0.96 s (WS1), 0.63 s (WS2), and 0.76 s (WS3), respectively.

Under the WS1 condition (Fig. 5(b)), the experimental data exhibit a moderate degree of dispersion even in the absence of deliberate wave excitation. This variability is likely attributable to residual micro-disturbances on the water surface, such as hydraulic circulation induced ripples within the indoor environment. Unlike the controlled laboratory setting, the pool environment lacks boundary damping, allowing even low-energy perturbations to persist. Furthermore, the geometric configuration of the setup - characterized by an incidence angle of approximately 88° - minimizes frequency-selective attenuation. At this steep angle, reflection occurs under quasi-specular conditions, with reflectivity approaching unity for both polarizations and exhibiting minimal sensitivity to frequency. As a result, the wave-induced attenuation remains largely invariant across the 120-230 GHz band.

As the water surface becomes more dynamic (Fig. 5(c) and (d)), corresponding to WS2 and WS3, respectively, the measured power loss distributions broaden significantly, particularly at the higher frequencies. This enhanced variability stems from the increasing irregularity of local surface slopes and the time-varying nature of the wavefront geometry, which introduces greater angular mismatch between the incident and reflected beams. At THz frequencies, the beam divergence is inherently narrow, and the angular alignment tolerance is small [41]. Therefore, even minor deviations from the specular reflection path due to wave-induced surface tilting can lead to significant degradation in received signal strength.

Despite the amplified dispersion, the central tendency of the data remains relatively stable across frequency. This consistency is a direct consequence of the grazing incidence configuration. Unlike the previous laboratory findings - where power loss increased with incidence angle due to the higher impact of surface roughness at grazing angles - the 88° angle employed here is well beyond the Brewster angle and closer to the regime of high reflectivity for both *v*- and *h*- polarizations. This geometry limits the role of polarization-dependent absorption and reduces the influence of frequency on reflection loss. As a result, the expected trend of increasing power loss with increasing incidence angle, observed in Fig. 2 for 45° and 60°, does not hold in this scenario and is instead suppressed due to the grazing propagation geometry.

The theoretical predictions, computed using the dual-scale I$^2$EM, are superimposed on the measured data as smooth red curves. At lower frequencies (120-160 GHz), the model aligns well with the central cluster of measurements, confirming its efficacy in capturing the average behavior of THz channel reflection over still or mildly dynamic surfaces. However, at higher frequencies (220-230 GHz), a discrepancy emerges between predicted and measured values. This deviation is primarily attributed to increased beam directivity and narrower beamwidth at higher frequencies, which exacerbate alignment sensitivity. In the presence of dynamic surface undulations, the beam's spatial footprint on the water surface becomes more susceptible to angular deflection and localized scattering, phenomena that are not fully captured by the idealized model assumptions. Additionally, practical imperfections such as minor tilt misalignments, or elevation vibration further widen the statistical spread at these frequencies.

*C. Power profile Performance*

To evaluate the generalizability of the statistical fading model under practical environmental conditions, a power profile analysis based on cumulative distribution functions (CDFs) was conducted for swimming pool measurements at 140 GHz and 220 GHz. The results, illustrated in Fig. 6, reveal consistent and interpretable trends that reinforce conclusions drawn from laboratory experiments above, while exposing new intricacies introduced by uncontrolled field conditions.

For both carrier frequencies, a clear monotonic degradation in SNR is observed with increasing wave amplitude. Specifically, under *calm surface conditions* (WS1), the SNR distribution is highly concentrated and steep, indicative of a stable and coherent propagation channel. Under *low-amplitude periodic waves* (WS2), the SNR distribution shifts rightward with broader dispersion, while *high-amplitude wave conditions* (WS3) yield the lowest and most widely spread SNR values. This trend is consisted with that in Fig. 3 and confirms again that water surface dynamics impose significant variability on the propagation channel path, primarily through perturbations in the reflection geometry and induced diffuse scattering. The consistency of this pattern across multiple frequencies highlights the dominance of surface-induced fading mechanisms over frequency-specific dielectric effects in such scenarios.

To quantify the fading behavior, Rayleigh, Rician, and Weibull models were fitted to the empirical CDFs. Table III presents the corresponding Fitting Gap and Fluctuation Error for each distribution under different wave conditions. The Weibull distribution consistently provides the best fit across all three wave states. In every case, it outperforms Rayleigh and Rician models, which assume more rigid fading structures. These results reinforce the premise that THz channel fading in aquatic NLOS environments is non-Gaussian and heavy-tailed, driven by a complex superposition of dynamic scattering events that conventional models fail to capture. The robustness of the Weibull distribution across both frequencies and wave states suggests its potential as a universal statistical descriptor for modeling THz power fluctuations over water surfaces.

To further dissect the fading structure, Fig. 7 shows the variation of Weibull scale and shape parameters as a function of frequency for WS1, WS2, and WS3. In contrast to laboratory observations - where both parameters exhibited predictable, monotonic trends with frequency and wave intensity - the swimming pool environment reveals a less regular and more scattered parameter evolution. This can be attributed to several field-specific factors. First, beam misalignment due to long-range propagation and equipment tilt becomes more pronounced at higher frequencies due to narrower beamwidths, resulting in greater susceptibility to angular jitter and power fluctuation. Second, the finite aperture of the receiver introduces sensitivity to micro-scale angular deflections caused by wavefront curvature, further complicating the interpretation of scale and shape parameters as smooth functions of frequency. This implies that static channel models are insufficient for system planning. Real-time channel state estimation, adaptive beam steering, and statistical learning techniques must be integrated to maintain reliable link performance.

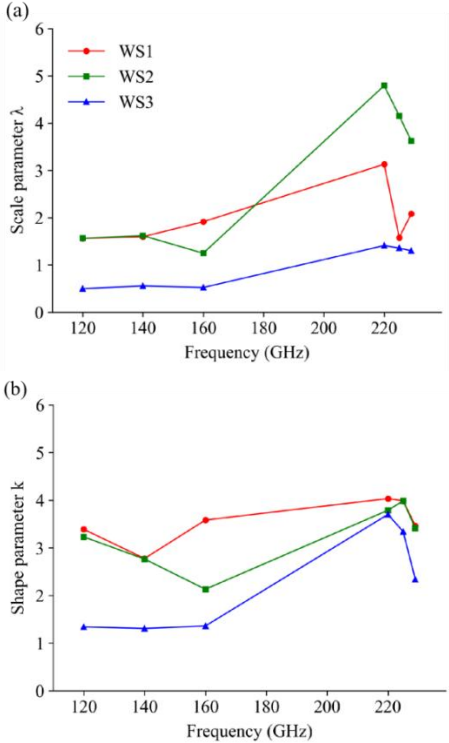

**Figure 7** Variation of Weibull parameters with frequency under different wave heights. (a) Scale parameter $\lambda$; (b) Shape parameter k. (b) keeps the same legend with (a).

*D. Channel performance prediction*

To evaluate the viability of reflection-assisted THz communication in real-world aquatic environments, we extended our modeling to the channel performance within a large-scale indoor venue - the Beijing National Aquatics Center, also known as the "Water Cube". With structural dimensions of 177 m × 177 m × 30 m, the facility presents an ideal setting for investigating NLOS THz channel propagation over reflective water surfaces. In such a configuration, the effective propagation distance - comprising the transmitter-to-water and water-to-receiver paths - is approximately 150 meters, and a shallow incidence angle is employed to mimic realistic deployment geometry in natatoriums.

The angular dependence of reflectivity plays a crucial role in determining the feasibility of such NLOS THz channels, especially under polarization-specific constraints. Fig. 8(a) presents the reflectivity profiles of 140 GHz and 220 GHz THz channels under both vertical (*v*-pol) and horizontal (*h*-pol) polarization. As expected, a sharp reflectivity null is observed for *v*-pol at the Brewster angle, validating fundamental electromagnetic boundary theory and aligning with earlier experimental observations in Fig. 2, where *v*-pol exhibited higher losses at 60° than at 45°. However, in the swimming pool scenario, where the incidence angle approaches 88°, the reflectivity behavior departs significantly from classical predictions. At such high incidence angles (corresponding to grazing angles), the interaction between the wavefront and water surface becomes highly sensitive to surface irregularities. The dominant factor affecting reflectivity is not angular scattering from increased projected roughness, but rather geometric shadowing caused by the surface wave slopes. As the incidence angle increases, more of the incoming wavefront becomes obstructed by the crests and troughs of surface waves, which blocks or deflects portions of the specular and scattering components. Consequently, both *v*-pol and *h*-pol channels experience reduced reflectivity - but due to the surface-induced shadowing (geometric shadowing). Based on these insights, we can define optimal angular ranges for efficient THz channel reflection. For the *v*-pol channel, effective reflectivity is achieved at incidence angles below 30°, and within a narrow high-angle band between 83° and 85°, where geometric effects begin to saturate. For the *h*-pol channel, which does not exhibit a Brewster minimum, the angular tolerance is broader - extending from approximately 50° to 85°. These angular bands are crucial for guiding the design of beam-steering protocols in future reflection-assisted THz communication systems, especially in multi-user, dynamically reconfigurable aquatic environments.

Path loss modeling was subsequently performed to quantify the total attenuation over a 150-meter channel, incorporating free-space loss, atmospheric absorption, and reflection loss from the water surface. The results, shown in Fig. 8(b), indicate multiple sharp absorption peaks due to atmospheric water vapor, particularly beyond 300 GHz, interspersed with relatively flat low-loss windows. Among them, 140 GHz and 220 GHz emerge as practical operational frequencies. Importantly, the simulations were conducted under an incidence angle of 85°, and parameterized using the reflection data from Fig. 8(a) and the fading characteristics from Fig. 6. While in-situ statistical measurements in the Water Cube are unavailable, this angle is sufficiently close to the 88° used in prior experiments, justifying the application of the same Weibull-based fading model. The results also confirm that h-pol waves exhibit slightly lower path loss than v-pol under this configuration, reinforcing its utility for high-reflection-angle deployments.

TABLE IV
BIT ERROR RATE CALCULATION PARAMETERS

| Parameter | Value |
|---|---|
| Transmit power | 0 ~ 40 dBm |
| Frequency | 140 GHz / 220 GHz |
| Angle of incidence | 85° |
| Distance | 150 m |
| Transmitting/Receiving antenna gain | 25 dBi |
| Transmitting/Receiving lens gain | 20 dBi |
| Noise power | -40 dBm |
| Water surface | WS3 |

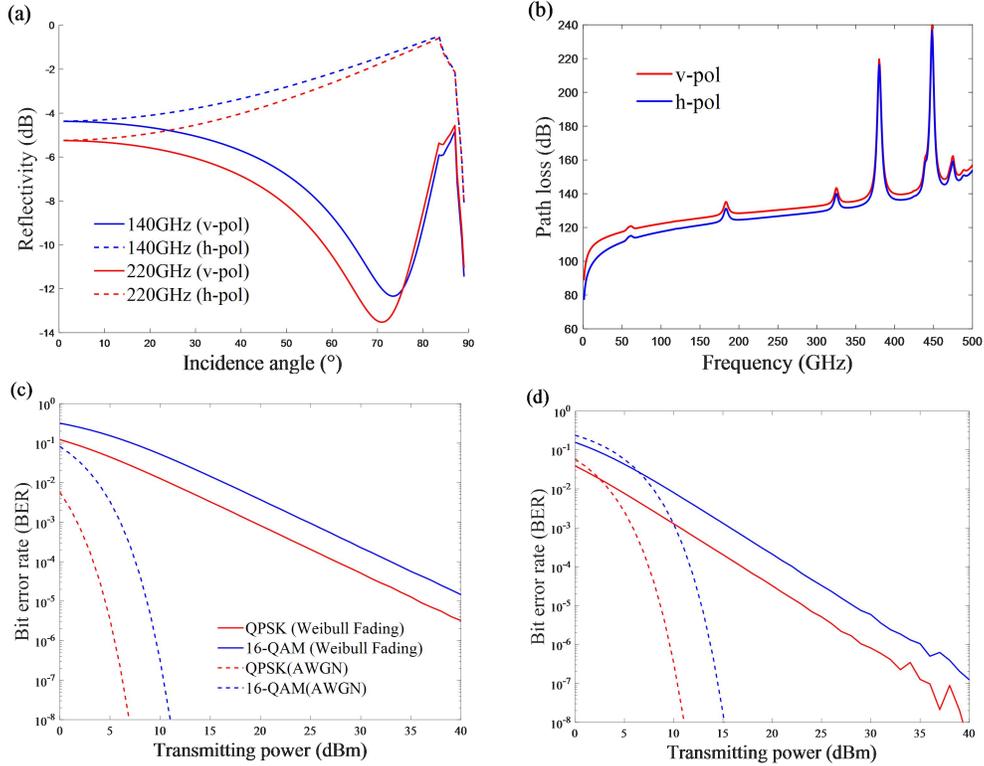

**Figure 8** Path loss and BER calculation. (a) Reflectivity at different incidence angles at 140 GHz and 220 GHz for different polarizations; (b) Total path loss for 150 m of signal transmission from 0 to 500 GHz; (c) BER calculation for 140 GHz channel with different modulations; (d) BER calculation for 220 GHz signal with different modulations.

To further assess system-level performance, we performed bit error rate (BER) simulations under realistic channel conditions. The parameters for this simulation, summarized in Table IV, include transmit power ranging from 0 to 40 dBm, a propagation distance of 150 meters, 25 dBi antenna gain, 20 dBi lens gain, and a noise power level of -40 dBm. The simulation is conducted using two commonly used modulation methods, namely quadrature phase-shift keying (QPSK) and 16-quadrature amplitude modulation (16-QAM). Under typical additive white Gaussian noise conditions, the theoretical BER calculation formulas for both are as follows:

$$\text{BER}_{\text{QPSK}} = \frac{1}{2}erfc\sqrt{\gamma} \qquad (8)$$

$$\text{BER}_{\text{16QAM}} = \frac{3}{8}erfc\sqrt{\frac{2}{5}\gamma} \qquad (9)$$

where $\gamma$ represents the SNR. However, in an environment with dynamic water surface reflection, the channel situation becomes more complex as the SNR no longer follows a simple probability distribution. Instead, we can characterize the BER through a characteristic function of the Weibull distribution. For QPSK under Weibull fading conditions, the BER will become

$$\text{BER}_{\text{QPSK}} = \int_0^\infty \frac{1}{2}erfc\sqrt{\gamma} \cdot f(\gamma)d\gamma \qquad (10)$$

where $f(\gamma)$ represents the probability density function of the

instantaneous SNR. It follows the following Weibull function:

$$f(\gamma) = \frac{k}{\lambda}\left(\frac{\gamma}{\lambda}\right)^{k-1} e^{-(\gamma/\lambda)^k} \quad (11)$$

where the specific values of the shape parameter $k$ and scale parameter $\lambda$ are obtained through fitting under WS3 conditions. The calculation of the BER for 16QAM follows a similar principle.

Fig. 8(c) and 8(d) display the BER curves at 140 GHz and 220 GHz, respectively, under Weibull fading for two modulation formats. As expected, QPSK significantly outperforms 16-QAM under all conditions, achieving lower BER at substantially lower transmit power. This can be attributed to its phase-only encoding, which offers robustness against amplitude fading and interference from angular scattering. On the other hand, 16-QAM's amplitude-and-phase modulation demands higher SNR, making it more vulnerable to multipath distortion and signal dispersion over rough surfaces.

A particularly noteworthy observation is the superior BER performance of the 220 GHz channel compared with its 140 GHz counterpart, even though both links operate with the same launch power. This finding might initially seem counterintuitive, as higher frequencies generally incur greater free-space path loss. Fitting the instantaneous SNR data with a two-parameter Weibull model reveals (see Fig. 7) that the shape parameter ($k$) rises from 1.4 at 140 GHz to 2.3 at 220 GHz. $k$ governs the thickness of the fading distribution's left-hand (deep-fade) tail [42]. A larger $k$ compresses this heavy tail and markedly lowers the probability of very low-SNR events, which implies a much lighter tail suffered by the 220 GHz, meaning far fewer excursions below the forward-error-correction threshold and, consequently, a lower BER. This highlights the importance of adaptive modulation and coding schemes in practical systems, where dynamically adjusting the modulation order and coding rate in response to fading can help sustain link reliability under such varying conditions.

## IV. CONCLUSION

In this article, we have systematically investigated THz channel performance affected by dynamic reflections from water surfaces, utilizing both theoretical modeling and experimental measurements. A modified dual-scale scattering model based on the Improved Integral Equation Model (I$^2$EM) was developed, explicitly accounting for the interplay between macroscopic sinusoidal waveforms and microscopic stochastic roughness on water surfaces. Laboratory experiments conducted over frequencies ranging from 120 GHz to 320 GHz, at incidence angles of 45º and 60º, confirmed the model's ability to accurately predict frequency- and angle-dependent channel power loss characteristics under varying wave conditions. Specifically, we observed stable power losses under calm surface conditions, which significantly increased and displayed higher variability as the water wave amplitude increased, due to enhanced surface scattering and angular dispersion.

Further real-world measurements in a natatorium environment validated the theoretical predictions under practical scenarios, including wave perturbations induced by swimmer activities. Through statistical analyses employing CDFs, it was demonstrated that the proposed model effectively characterizes the stochastic nature of channel variations, with superior fitting accuracy compared to conventional models. Moreover, BER simulations highlighted a marked degradation in communication performance under realistic wave disturbances, underscoring the necessity of employing adaptive modulation and coding schemes. These findings collectively provide critical insights and practical guidance for future design and deployment of robust THz communication systems in complex aquatic environments.